\documentclass[hyper]{JHEP} 

\usepackage{epsfig}

\newcommand\fverb{\setbox\pippobox=\hbox\bgroup\verb}
\newcommand\fverbdo{\egroup\medskip\noindent%
			\fbox{\unhbox\pippobox}\ }
\newcommand\fverbit{\egroup\item[\fbox{\unhbox\pippobox}]}
\newbox\pippobox

\title{Matrix model and string field theory}

\author{by J. Kluso\v{n}\\

	 Department of Theoretical Physics and 
Astrophysics\\

                   Faculty of Science, Masaryk University\\

Kotl\'{a}\v{r}sk\'{a} 2, 611 37, Brno\\

Czech Republic\\

	E-mail: \email{klu@physics.muni.cz}}

\preprint{\hepth{0011029}}	

\abstract{In this short note we would like to 

show the relation between 
the cubic open string field theory for $N$ D-instantons
and the string field theory  in the presence of 
the background  B field.}

\keywords{D-branes, string field theory, noncommutative geometry}

\def\tr{\mathrm{Tr}}

\def\bra #1{\left<#1\right|}

\def\ket 
#1{\left|#1\right>}
\begin{document}



\section{Introduction}
Tachyon condensation has being one of the
most studding problems in  string theory
in the past two years
\cite{SenP,witen,Horava},  for review see
\cite{Olsen, Lerda,Schwarz} and for the recent discussion
the relation between the tachyon condensation and
K-theory, see \cite{witen2,Matsuo,Moore}.
Evidence for this proposal was given from the
analysis of CFT description of this system \cite{SenP},
for review of this approach, see \cite{Lerda,
Schwarz}. It was also shown on many examples that
string field theory approach to this problem is very
effective in the calculation of the tachyon potential
\cite{WittenSFT,SenFT1,SenFT2,BerkovitzFT1,HarveyFT,SenFT3,
TaylorFT,KochFT,Desmet,NaqviFT,SenFT4,DavidFT,WittenFT,
RasteliFT,SenFT5,TaylorFT2,KochFT2,NaqviFT2,Schnabl,SenN1,Hata,
SenN2,TaylorN,Feng},
for review see \cite{Ohmori}.
This problem was recently studied from the point of view 
of the Witten's background independent open string
field theory \cite{WittenBT,WittenBT1,WittenBT2,ShatasviliBT,
ShatasviliBT1,ShatasviliBT2,MooreBT,SenBT,Cornalba,Okuyama,MooreBT2,
DasguptaBT1,ShatasviliBT3,KrausBT} as well.
Success of the string field theory in the analysis of 
the tachyon condensation indicates that the string field theory could play
more fundamental role in the nonperturbative formulation of string theory.

The second approach to the problem of tachyon condensation
is based on the idea of  the noncommutative geometry \cite{WittenNG}.
This analysis has been inspired with the seminal paper
\cite{StromingerA}. Application of  this approach to the problem
of the tachyon condensation was pioneered in
\cite{Gopakumar,Harvey}. This research was then developed
in other papers \cite{Harvey2,Strominger2,Rey,Matsuo,
SenTP,Mukhi,Strominger3,HarveyS,Larsen}. The tachyon condensation
has been also studied from the point of view
of the matrix model \cite{KlusonM,Kraus,LiT}.
Success of these two approaches immediately leads to
the question whether there is any relation between the
matrix model and consequently noncommutative theory 
\cite{Seiberg} and the string field theory. The similar
problem was analysed in the recent paper \cite{HarveyS} in
the framework of the effective string field theory.

In this paper we would like to address this problem  
in the context of the string field theory for general
configuration of $N$ D(-1)-branes (D-instantons) 
in bosonic string theory. We propose the generalised form
of the open string field theory \cite{WittenSFT} that
allows description of  any configuration of D-instantons.
We will make many consisitent checks justifiing our
approach, in particular, we will show that the new
matrix valuded BRST operator is nilpotent on condition
that the background configuration of $N$ D-instantons
obeys the equation of motion familar from the matrix
models \cite{BanksM,Banks,Li,Aoki,
Ishibashi,Ishibashi1}. We will show that this theory
obeys generalised form of the string field theory
axioms that were recently discussed in the
papers \cite{Lazaraiu1,Lazaraiu2,Lazaraiu3}.
We will also show that the emergence of
lower dimensional D-branes from D25-brane is
very natural proces from the point of view of
matrix string field theory. In fact, the efficency of
the matrix theory description of the tachyon condensation
was recently stressed in \cite{Kraus,LiT,Mandal,Alwis}.

In section (\ref{second}) we review the basis facts
about cubic string field theory \cite{WittenSFT}. We
will be very brief, more information can be found in
the nice papers \cite{SenFT1,SenFT2,TaylorFT,Ohmori}.

In section (\ref{third}) we will discuss the
string field theory for $N$ D(-1)-branes (D-instantons). We propose
the modification of the BRST operator for $N$ D-instantons
in such a way that we will be able to take into
account their  general space-time positions.
We will show that when we consider the  
configuration of $N$ separate D-instantons then
 off-diagonal modes of string field become massive
according to the nonzero string winding charge.
 Then we will study the noncommutative
background of D-instantons and we  show that
the string field action for $N$ D-instantons in the limit 
$N\rightarrow \infty$ becomes the string field theory
action for D-brane with the noncommutative world-volume.
In this way we will show that  string field theory
of $N$ D-instantons is able to describe all even dimensional D-branes in
the same way as  D-branes emerge in  
the matrix theory \cite{Seiberg,BanksM,Banks,Li,Aoki,
Ishibashi,Ishibashi1}.

In section (\ref{fourth}) we will discuss the tachyon
condensation on the world-volume of the D25-brane
in the presence of the background B field. Using the
result given in the section (\ref{third}) we present
a simple solution describing the tachyon condensation
into $k$ D2p-branes.

In the conclusion (\ref{sixth}) we will discuss some open
problems and implication of our results.




\section{Brief review of string field theory}\label{second}

In this section we will briefly review the Witten's open
string field theory \cite{WittenSFT}. The Witten's
formulation is in noncommutative extension of
differential geometry, where string fields, BRST operator
$Q$ and the integration over string configuration
$\int$ in string theory are analogies of differential 
forms, the exterior derivative   $d$ and the integration
over the manifold $M$ in the differential geometry,
respectively. The ghost number assigned to
the string field corresponds to the degree of differential
form. Also the noncommutative products between
string fields $\star$ is interpreted as an analogy
of the wedge product $\wedge $ between differential forms. 

The axioms obeyed the system of $\int, \star $ and
$ Q$, are
\\
\begin{eqnarray}\label{ax}
\int QA=0 \ , \nonumber \\
Q(A\star B)=(QA)\star B+(-1)^{A} A\star (QB)\ , 
\nonumber \\
(A\star B)\star C=A\star (B\star C) \ , \nonumber \\
\int A\star B=(-1)^{A B}\int B\star A \ , \nonumber \\
\end{eqnarray}
where $A,B,C$ are arbitrary string fields. As was shown
in \cite{WittenSFT} in order to describe a gauge invariant
string field theory we must include the full Hilbert space
of states of the first quantized open string theory including
the $b$ and $c$ ghost fields, subject to the condition that
the state must  carry ghost number $1$. Here we are using
the convention that $b$ carries ghost number $-1$ and
$c$ carries ghost number $1$ and $SL(2,R)$ invariant vacuum
$\ket{0}$ carries ghost number $0$. 
In the previous expression $Q$ is BRST operator of the
first quantized open string. 
The string field theory action for Dp-brane is  
\footnote{We use normalisation given in \cite{SenFT1}
and we work in Euclidean signature.}
\begin{equation}\label{Waction}
S=\frac{2\pi^2T_p}{g_s}
\int \left(\frac{1}{2}\Psi\star Q\Psi +
\frac{1}{3}\Psi\star\Psi\star\Psi
\right) \ ,
\end{equation}
where $T_p=2\pi/(4\pi^2\alpha')^{(p+1)/2}$ is a D-brane tension
and $g_s$ is a string coupling constant.
This abstract form of the string field action can be written in
the other form appropriate  for the calculation. 
It is useful to write it  in terms of the conformal field theory (CFT)
\cite{LeClair}. Let $\ket{\Phi}$ be an arbitrary state
in $\mathcal{H}$ the full Hilbert space of states
of the first quantized open string theory and let
$\Phi(x)$ be a local field (vertex operator) in conformal
field theory which creates this state $\ket{\Phi}$ from
out of $SL(2,R)$ invariant  vacuum
\begin{equation}
\ket{\Phi}=\Phi(0)\ket{0} \ .
\end{equation}
In the CFT language, the string field action is given 
\begin{equation}
S=\frac{2\pi^2 T_{p}}{g_s}
\left(\frac{1}{2}\bra{\Psi}Q\ket{\Psi}+
\frac{1}{3}\left<f_1\circ \Phi(0) f_2
\circ \Phi(0) f_3\circ \Phi(0) \right>\right) \ ,
\end{equation}
where $f_i$ are known conformal maps reviewed
in \cite{SenFT1} and $f\circ \Phi(0)$ denotes
conformal transformation of the vertex operator $\Phi$ by $f$.

We must also mention that there is a formulation of
the string field theory in terms of the operator formalism
\cite{Gross,Samuel}. The operator formalism was used
in \cite{Sugino,Kawano} where the string field theory
in the constant B field background was studied.

In the following we will work mainly with the abstract
definition of the string field theory given in \cite{WittenSFT}.
In the next section we will discuss the string field theory
for $N$ D-instantons.

\section{String field theory for $N$ D-instantons}\label{third}

In this section we propose the action for $N$ D-instantons. This
can be done very easily in such a way that all string
fields will carry  the  indeces corresponding to the adjoint 
representation of the gauge group $U(N)$.
Then the string field action  for $N$ D-instantons has a form
\begin{equation}
S=\frac{2\pi^2T_{-1}}{g_s}\int \left(\frac{1}{2} \Psi_{ij}\star \tilde{Q}^{inst}_{jk}
\Psi_{ki}+\frac{1}{3}\Psi_{ij}\star\Psi_{jk}\star\Psi_{ki} \right) \ ,
\end{equation}
where 
\begin{equation}
\tilde{Q}_{ij}^{inst}=\tilde{Q}^{inst} \delta_{ij} \ ,
\end{equation}
with the BRST operator $\tilde{Q}^{inst}$ of the string living on
one single D-instanton. However,
there is a one important issue with this action. This action describes
the fluctuations around the background corresponding to 
$N$ D-instantons in the same place in the space-time so that
the $U(N)$  symmetry of the action is unbroken. Under this symmetry the
string field transform as $\Psi'=U\Psi U^{-1} , U\in U(N)$.
 In order to include
the more general background configuration of $N$ D-instantons,
we propose the new BRST operator for $N$ D-instantons in
the form
\begin{equation}\label{Qinst}
Q_{ij}=Q^{inst}_{ij}+Q^0_{ij}, \
\end{equation}
where $Q^{inst}_{ij}=Q^{inst}\delta_{ij}$
 is the instanton BRST operator  without
zero mode part and $Q^0_{ij}$ is a generalised zero mode
part of the BRST operator for $N$ instantons in the form
\begin{eqnarray}
Q^0_{ij}=\sum_{n=-\infty}^{\infty} c_n L_{-n}^0=
\frac{1}{2}c_0
g_{IJ}(\alpha^I_0\alpha^J_0)_{ij}+
\sum_{n=-\infty, n\neq 0}^{\infty}c_n g_{IJ}(\alpha_{-n}^I\alpha_{0}^J)_{ij}\ ,
\nonumber \\
\frac{1}{2}c_0g_{IJ}(\alpha_0^I\alpha^J_0)_{ij}=\frac{1}{4\pi^2\alpha'}
c_0g_{IJ}[X^I,[X^J, \cdot \  ]]_{ij} \ ,
c_ng_{IJ}(\alpha_{-n}^I\alpha_0^J)_{ij}=\frac{\sqrt{2}}{2\pi\sqrt{\alpha'}}
c_ng_{IJ}(\alpha_{-n}^I[X^J,\cdot \ ])_{ij} \ , \nonumber \\
\end{eqnarray}
 where 
the action of this operator  on 
any string field $\Psi$ is defined  as
\begin{equation}\label{Q0act}
Q(\Psi)_{ij}=Q^{inst}\Psi_{ij}+c_0\frac{1}{4\pi^2\alpha'}
g_{IJ}[X^I,[X^J,\Psi]]_{ij}+\frac{\sqrt{2}}{2\pi\sqrt{\alpha'}}
\sum_{n=-\infty,n\neq 0}^{\infty}
 c_ng_{IJ}\alpha_{-n}^I[X^J,\Psi]_{ij} \ .
\end{equation}
and where $X^I, I=1,\dots, 26$ are $N\times N$ matrices
describing the background configuration of $N$ D-instantons.
And finally, the normalisation of the various terms given above
will be clear from next discussion.

Since (\ref{Qinst}) differes from the ordianary BRST
operator, in particular, it is matrix valued, we should prove that
it is nilpotent and that the string field theory defined in this way
obeys all string field theory  axioms  (\ref{ax}).

We start with the proof of the nilpotence of $Q$ whose
obvious generalisation  is
\footnote{When it is not explicitly written, expressions with
the same indeces will correspond to the sumation over them.}
\begin{equation}
Q_{ij}Q_{jk}=0 \ .
\end{equation} 
Since we know
that all new properties are included in the zero mode part
of $\alpha_0$, in particular, the ghost part is the same as in
the abelian case, it is sufficient for our purposes to show 
that $L_n$  obye the correct 
Virasoro algebra.
 In order to do that we define oscilator
modes $\alpha_m$ as follows
\begin{equation}
(\alpha_m^I)_{ij}=\alpha_m^I\otimes \delta_{ij}, \ m\neq 0 , \
[(\alpha_m^I),(\alpha_n^J)]_{ij}=
m\delta_{m+n}\delta^{IJ}\delta_{ij} \ , m, n \neq 0 \ ,
\end{equation}
where the matrix multiplication is undestood.
It is also clear that $\alpha_0^I$ commutes with $\alpha_m^J$ since
$\alpha_m^I, m\neq 0$ are proportional to the identity matrix in the
space of the Chan-Paton factors and $X_{ij}$ commutes with $\alpha_m$
from the basis definition of the commutation relations. The only
notrivial task is to compute the commutator 
$[\alpha_0^I,\alpha_0^J]$.
Firstly we define Virasoro generators
\begin{equation}
(L_m)_{ij}=\frac{1}{2}\sum_{n=-\infty}^{\infty}g_{IJ}(\alpha^I_{m-n}
\alpha^J_{n})_{ij} \ ,
(L_0)_{ij}=\frac{1}{2}g_{IJ}
(\alpha_0^I\alpha_0^J)_{ij}+g_{IJ}\sum_{n=1}^{\infty}
(\alpha_{-n}^I\alpha_n^J)_{ij} \ 
\end{equation}
with
\begin{equation}
(\alpha_0^I)_{ij}=\frac{1}{\pi
\sqrt{2\alpha'}}[X^I,\cdot \  ]_{ij} \ .
\end{equation}
Now the commutator of two $\alpha_0^I$ is
(when acts on any matrix $M$) equal to
\begin{eqnarray}\label{0com}
[\alpha_0^I,\alpha_0^J]M=
\frac{1}{\sqrt{2\alpha'}\pi}
(\alpha_0^I[X^J,M]-\alpha_0^J[X^I,M])=\nonumber \\
=\frac{1}{2\pi^2\alpha'}
([X^I,[X^J,M]]-[X^J,[X^I,M]])= 
\frac{1}{2\pi^2\alpha'}[[X^I,X^J],M] \ .  \nonumber \\
\end{eqnarray}
Now we are ready to calculate the commutator $[L_m,L_n]$.
In fact, the calculation of this commutator is
well known for a long time, see for example \cite{WittenBook}.
 The novelty in our
approach is in the presence of the matrix valued zero mode
operators $\alpha_0$. 
To ilustrute this issue let us work out the commutator that
is present in the calculation of the commutator of two
Virasoro generators with $m+n\neq 0$
\begin{eqnarray}
[g_{IJ}\alpha^I_m\alpha^J_0,g_{KL}
\alpha^K_n\alpha_0^L]
=g_{IJ}g_{KL}(\alpha_m^I\alpha^K_n\alpha^J_0
\alpha^L_0-\alpha_n^K\alpha^I_m\alpha^L_0\alpha^J_0)
=\nonumber \\
=g_{IJ}g_{KL}(g^{IK}m\delta_{m+n}\alpha^J_0
\alpha^L_0+\alpha^K_n\alpha^I_m
[\alpha_0^J,\alpha^L_0])
=g_{IJ}g_{KL}\alpha^K_n\alpha^I_m
[\alpha^J_0,\alpha_0^L] \ , \nonumber \\
\end{eqnarray}
where we have used the fact that $m+n\neq 0$.
It can be shown that with nonzero upper result
we cannot obtain the correct form 
 of the Virasoro algebra. For that reason
we must demand the vanishing of the
 commutator $[\alpha_0^I,\alpha_0^J]$
that leads to the condition
\begin{equation}\label{conformal}
[X^I,X^J]=i\theta^{IJ}1_{N\times N}
\end{equation}
as we can see from (\ref{0com}). It is
clear that the commutator can be nonzero
only in the case of
 infinite dimensional matrices. 
Note that this expression has a form of
the  solution of the
equations of motion obtained from the matrix
model \cite{BanksM,Banks,Li,Aoki,Ishibashi,Ishibashi1}
\begin{equation}
[X^I,[X^I,X^J]]=0 \ .
\end{equation}
This result is an analogue of the case of
the string propagating in the general background when
the requirament of the conformal invariance leads
to the condition that the background fields should be
solutions of the equation of motion obtained from the
space-time effective action. In our case this effective
action is D-instanton effective action \cite{Li,Aoki} 
\begin{equation}
S\sim \tr g_{IK}g_{JL}[X^I,X^J][X^K,X^L] \ .
\end{equation}
 Using (\ref{conformal}) it is now straightforward to
prove that the Virasoro algebra has a correct form.
We do not repeat the standart analysis here, more
details can be found in \cite{WittenBook} where it was 
also shown how we can determine the central charge
of the Virasoro algebra so that we obtain the result
\begin{equation}
[L_m,L_n]_{ij}=(m-n)L_{ij}+A(m)\delta_{m+n}\delta_{ij},
\ i,j=1,\dots, N \ ,
\end{equation}
where $A(m)=\frac{1}{12}c(m^3-m)$ is a central charge of the
Virasoro algebra. In the previous expression we have written 
explicitelly the matrix indeces $i,j$ to stress the matrix
nature of the generators $L_m$. Some comments
about the previous result. Since the central charge
 is not affected by matrix
valued nature of the Virasaro generators
it  is proportional to the unit matrix. In the same way we can
argue that the ghost part of the action does not depend on
the matrix notation. Then it immediatelly follows  that the
generalised BRST operator is nilpotent
\begin{equation}
Q^2_{ij}=\frac{1}{2}\{Q,Q\}_{ij}=0 \ ,
\end{equation}
in case of the critical bosonic string theory $D=26$  \cite{WittenBook}.
Of course, for the existence of the nilpotent BRST generator
is cruical the condition (\ref{conformal}) which is nothing
else than the requirament that
the background configuration of $N$ D-instantons must be
solution of the equations of motion of the low energy action.

Now we are ready to prove that the string field theory with
the  generalised
BRST operator (\ref{Qinst}) obeys all axioms given in (\ref{ax}).
In fact, we should consider the more general form
of  the axiomatic formulation  which is appropriate for the
 general configuration
of D-branes. In fact, this has been done in the abstract
form in \cite{ZwiebachAB1,ZwiebachAB2} and more recently
in the series of papers \cite{Lazaraiu1,Lazaraiu2,Lazaraiu3}.
We do not mean to discuss these general constructions, see
for example \cite{Lazaraiu2} for very nice explanation. For
our purposes it is sufficient to know that now the BRST operator
$Q$ depends on the background configuration of D-branes.
For that reason we should generalise the axioms given
in (\ref{ax}) in order to include these properties. Intuitevly,
this can be seen as follows. Let us presume that we have a
configuration of $N$ D-instantons described with the matrices
\begin{equation}\label{clasI}
X^I=\left(\begin{array}{cccc}
x^I_1 & 0 &\dots & 0\\
0 & x^I_2 & \dots & 0 \\
\dots & \dots & \dots & \dots \\
0 & \dots & \dots & x^I_N \\ \end{array}\right), \ I=1,\dots,26 \ ,
\end{equation}
then the second term in (\ref{Q0act})
acting on any string field $\Psi_{ij}$ gives
\begin{eqnarray}
g_{IJ}[X^I,[X^J,\Psi]]_{ij}=g_{IJ}[X^I_{im}[X^J,\Psi]_{mj}-
[X^J,\Psi]_{im}X^I_{mj}]=\nonumber \\
=[x^I_i\delta_{im}[x^J_m\delta_{mk}\Psi_{kj}-\Psi_{mk}
x^J_k\delta_{kj}]-[x^J_i\delta_{ik}\Psi_{km}-\Psi_{ik}\delta_{km}x^J_k]
x^I_m\delta_{mj}]g_{IJ}=\nonumber \\
=[x^I_i\delta_{im}[x^J_m-x^J_j]\Psi_{mj}-[x^J_i-
x^J_m]\Psi_{im}x^I_m\delta_{mj}]g_{IJ}=\nonumber \\
=g_{IJ}[x^I_i-x^I_j][x^J_i-x^J_j]\Psi_{ij} \ , \nonumber \\
\end{eqnarray}
which, after multiplication with $\frac{1}{2(2\pi\alpha')^2}$,
gives precisely the value
$\frac{1}{2}(M^2_{ij})$ where $M_{ij}^2=\frac{(\bigtriangleup
x)^2_{ij}}{(2\pi\alpha')^2}$ is a minimal mass for
 the string stretched between i-th
and  j-th D-instanton.  Of course, it is slightly awkward to 
speak about the mass of the string with all Dirichlet conditions but
as we will see in the moment the D-instanton configuration
allows also existence of higher dimensional D-brane and
then the previous calculation could be applied to the string stretched
between different D-branes so that this analysis is correct.
 In the
same way we can analyse the third term in (\ref{Q0act}).
Finally,  the emergence of the second
factor $\alpha'$ in the previous result  comes from
 $\alpha'p^2$ in the BRST operator. This
momentum operator naturally arises from particular
configuration of $N$ instantons as we will show in a moment. 
The previous result suggests that for D-instantons background configuration
 (\ref{clasI}) the generalised BRST operator
$Q$ can be written as a sum of the BRST operators
\begin{equation}\label{Ninst}
Q_N=\sum_{i,j}Q_{ij} \ ,
\end{equation}
where $Q_{ij}$ is the BRST operator for the string connecting
i-th and j-th D-instanton. It is uderestood in
the previous expression that $Q_{ij}$ acts on the string field
sector corresponding to the string going from i-th
to j-th D-instanton only and consequently 
the BRST operator $Q_{ij}$ 
 does not act on  string states from different
string sectors.  It is well known that for each $i,j$ string
sector the string field theory is correctly defined.
Then we can very easily generalise the axioms (\ref{ax})
for any D-instanton background in the same way as in
\cite{Lazaraiu1,Lazaraiu2,Lazaraiu3}.  Firstly, it
is clear that the generalisaton of the expression $\int \Psi$ which
is linear in CP indeces is given as $\int \tr \Psi$. With this
definiton it is natural the generalise the expression
$\int Q \Psi$
as follows
\begin{equation}\label{nax1}
\tr \int Q\Psi=0
\end{equation}
which for (\ref{clasI}) gives
\begin{equation}\label{ax1}
 \int Q_{ij} \Psi_{ji}=\sum_{ij}\int Q_{ij}\Psi_{ij}=0 \ .
\end{equation}
We then see that the matrix valued BRST operator
obeys the generalised first axiom in (\ref{ax}) for the
 background (\ref{clasI}). Since we know that
$Q$ is the correct BRST operator for any configuration of D-instantons
(obeying (\ref{conformal})) and as is well
known from the matrix theory proposal there is not any fundamental
difference between (\ref{clasI}) and more general 
configurations given in (\ref{conformal})
 we can claim that the generalised $Q$
operator obeys (\ref{nax1}) as well.

As a next thing we turn to the second axiom in (\ref{ax}).
As was shown in \cite{Lazaraiu2} this axiom should be modified
in the presence of D-branes as follows. Let $A_{ij}$ corresponds
to some string field for the string 
stretching between i-th and j-th D-instanton. Then it is clear
that this string can be glued with strings ending or starting
on i-th or j-th D-instantons. We see that the gluing operation $\star$
is naturally generalised to the matrix valued multiplication between
matrix valued string fields $A, B$. We then obtain
the string field$(A\star B)_{ij}=A_{ik}\star B_{kj}$ 
which corresponds to the string going from i-th to j-th D-brane
that arises from the string stretching between i-th D-instanton and k-th D-instaton
where it glues with the string stretching between 
k-th D-instanton and j-th instanton. Since there is no prefered D-instanton we should
sum over all D-instantons which corresponds to the sum over $k$ in
the upper expression. 
It is then natural to expect that the appropriate BRST operator
acting on this string (going from i-th to j-th D-instanton)
 is $Q_{ij}$. Then we immediatelly obtain
the generalisation of the second axiom in (\ref{ax}) in the
form \cite{Lazaraiu2}(No summation over $i,j$ and we
explicitly write the sum over $k$.)
\begin{equation}\label{axd2}
Q_{ij}(A\star B)_{ij}=\sum_k (Q_{ik}A_{ik})\star B_{kj}
+(-1)^{A}\sum_k A_{ik}\star (Q_{kj}B_{kj}) \ .
\end{equation}
We propose  more general form of the second axiom
in (\ref{ax}) that reduces to (\ref{axd2}) in case of 
D-instanton background (\ref{clasI}). This new form has
an andventage that holds for any configuration of D-instantons
on condition of validity (\ref{conformal}). Simply, we propose
that the second axiom in (\ref{ax}) is 
\begin{equation}\label{nax2}
Q(A\star B)_{ij}=(QA)_{ik}\star B_{kj}+(-1)^{A} A_{ik}\star
(QB_{kj}) \ ,
\end{equation}
where the acting of $Q$ on string fields is defined in
(\ref{Q0act}).
For (\ref{clasI}), the left hand side of (\ref{nax2}) is 
equal to
\begin{equation}
Q(A\star B)_{ij}= Q_{ij}(A\star B)_{ij}
\end{equation}
and the right hand side  
\begin{equation}
(Q(A)\star B)_{ij}+(-1)^{A}(A\star Q(B))_{ij}=
\sum_kQ_{ik}(A)_{ik}\star B_{kj}+
\sum_k (-1)^{A}A_{ik}\star (Q_{kj}B_{kj}) \ .
\end{equation}
  so that we obtained from (\ref{nax2}) 
 the generalised second axiom
(\ref{axd2}). Using these results we can claim that the
matrix valued BRST operator $Q$ obeys the first  two
generalised axioms (\ref{nax1}),(\ref{nax2}).
To see this more preciselly, we can argue as follows.
 It is natural
to expect that general configuration of D-instantons (when
the background obeys the matrix theory equation of motion)
arises from (\ref{clasI}) as its solution
of equation of motion. In other words, let us presume  that general
BRST operator can be written as
\begin{equation}
QA=Q_NA+\Phi_0\star A-(-1)^AA\star \Phi_0 \ ,
\end{equation}
where $A$ is any string field and where the matrix multiplication
is underestood (We will say more about this approach in
the next section). From the fact that $Q^2=Q_N^2=0$
we  get from the upper expression (No summation over $i,j$)
\begin{eqnarray}
(Q^2A)_{ij}=0=
\left[Q_{ik}(\Phi_0)_{ik}+
(\Phi_0)_{il}\star(\Phi_0)_{lk}\right]\star A_{kj}+\nonumber \\
+(-1)^{2A+1}A_{ik}\star\left[
Q_{kj}(\Phi_0)_{kj}+(\Phi_0)_{kl}\star(\Phi_0)_{lj}\right]\nonumber \\
\end{eqnarray}
We see that the  general BRST operator will be nilpotent
in case when the string field $\Phi_0$ obeys the 
string field equation of motion for D-instntons background 
(\ref{clasI}) (No summation over $i,k$)
\begin{equation}
Q_{ik}(\Phi_0)_{ik}+(\Phi_0)_{il}\star(\Phi_0)_{lk}=0 \ .
\end{equation}
Then it is easy to see that the  generalised BRST operator
obeys all axioms given in (\ref{ax}). Firstly, we have
\begin{equation}
\tr\int QA=\int \left(\sum_{i,j}Q_{ij}A_{ij}
+(\Phi_0)_{ij}\star A_{ji}-(-1)^AA_{ij}\star (\Phi_0)_{ji}\right)=0 \ ,
\end{equation}
where we have used the fact that $Q_{ij}$ obeys the  first axiom
in (\ref{ax}) and also we have used the fourth axiom in
(\ref{ax}) togehter with the fact that $\Phi_0$  has a ghost
number one. 
We can also show that (No summation over $i,j$)
\begin{eqnarray}
Q(A\star B)_{ij}=Q_{ij}(A\star B)_{ij}+(\Phi_0)_{ik}
\star (A\star B)_{kj}-
(-1)^{A+B}(A\star B)_{ik}\star(\Phi_0)_{kj}=\nonumber \\
=(QA)_{ik}\star B_{kj}+(-1)^AA_{ik}\star (QB)_{kj} \ .
\nonumber \\
\end{eqnarray}
So we see that the  the generalised BRST operator obeys
the second axiom  (\ref{nax2}).

As a futher support of our proposal we will consider
the background configuration of D-isntantons
in the form 
\begin{equation}\label{non}
[X^a,X^b]=i\theta^{ab}1_{N\times N}
, \ a,b=1,\dots ,  2p, \ X^i=0, \
i=2p+1,\dots, 26 \ .
\end{equation}
We see that (\ref{non}) belong to the class
of the background configuration (\ref{conformal})
hence BRST operator (\ref{Qinst}) is nilpotent and defines
 correct
string field theory. 
Then we get
\begin{eqnarray}\label{instder}
\frac{1}{4\pi^2\alpha'}g_{ab}[X^a,[X^b,\Psi]]=
\frac{1}{4\pi^2\alpha'}g_{ab}\theta^{ac}\theta^{bd}
\theta_{ce}\theta_{df}[X^e,[X^f,\Psi]]=\nonumber \\
=-\alpha' G^{ab}[C_a,[C_b,\Psi]]  \ ,
\nonumber \\
\end{eqnarray}
where we have used 
\footnote{We work  with the
metric in the diagonal form $g_{IJ}=g_{II}\delta_{IJ}$.}
\begin{equation}\label{opp}
G^{ab}=-(2\pi\alpha')^{-2}\theta^{ac}
g_{cd}\theta^{db}, \ C_a=-i\theta_{ab}X^b \ ,
\end{equation}
In the same way we obtain
\begin{eqnarray}\label{Non}
\frac{\sqrt{2}}{2\pi\sqrt{\alpha'}}c_ng_{ab}
\alpha_{-n}^a[X^b,\Psi]=
\frac{i\sqrt{2}}{2\pi\sqrt{\alpha'}}c_n(2\pi\alpha')^2
G^{ce}\theta_{cd}\alpha^d_{-n}[C_e,\Psi]=\nonumber \\
=-i\sqrt{2\alpha'}c_nG^{ab}\tilde{\alpha}_{a,-n}
[C_b,\Psi] \ , \ \tilde{\alpha}_{a,-n}=
-(2\pi\alpha')\theta_{ab}\alpha^b_{-n} \ .\nonumber \\
\end{eqnarray}
We see that the expression
$[C_a,\Psi]_{ij}$ is the derivation in the operator 
formalism of the noncommutative theory
\cite{StromingerA,Strominger2}. 
Using the correspondence between operators
and  functions on noncommutative space-time we
obtain
\begin{equation}
[C_a,\Psi] \Leftrightarrow \partial_a \Psi(x) \ ,
\end{equation}
where now $\Psi$ looses all the gauge group indeces
and becomes function on the noncommutative 
space-time with coordinates $x^a, a=1,\dots,2p$.

This can be seen more precisely as follows. The 
general string field \cite{WittenSFT} should have a
ghost number $1$. For that reason we can write
any string field as 
\begin{equation}
\ket{\Psi}_{ij}=\sum_{n,m,l} A_{ij}^{nml}
\ket{n,m,l} \ ,
\end{equation}
where
\begin{equation}
\ket{n,m,l}=
\alpha_{-n_1}^{\mu_1}\dots \alpha^{\mu_i}_{-n_i}
b_{-m_1}\dots b_{-m_j}c_{-l_1}\dots c_{-l_k}|
\Omega \rangle , \ |\Omega\rangle=c_1\ket{0}\ ,
n>0,m>0,l \geq 0
\end{equation}
form the basis of the of the Hilbert space of
the first quantised open string restricted to
the states with the ghost number $1$ and obeying 
Dirichlet boundary conditions. Note that
$A^{nml}$ are $N\times N$ matrices
describing various string fields. In the following we
will presume the limit $N\rightarrow \infty$.

From (\ref{non}) we can also anticipate that the 
correct open string quantities are \cite{Seiberg}
\begin{eqnarray}\label{OPM}
G_{ab}=-(2\pi\alpha')^2\left(Bg^{-1}B\right)_{ab}, \ \nonumber \\
G_s=g_s\det(2\pi\alpha' Bg^{-1})^{1/2}, \ \nonumber \\
B_{ab}=\left(\frac{1}{\theta}\right)_{ab} \ , \nonumber \\
\end{eqnarray}
and consequently
\begin{equation}
\frac{T_{-1}}{g_s}=\frac{T_{2p-1}}{G_s} 
\sqrt{\det G}(2\pi)^p\sqrt{\det \theta} \ .
\end{equation}
Then we can write 
\begin{eqnarray}\label{e}
\frac{2\pi^2 T_{-1}}{g_s}\tr\bra{\Psi}\frac{-\alpha'}{2}
G^{ab}c_0[C_a,[C_b,\ket{\Psi}]]
=-\frac{2\pi^2 T_{2p-1}}{G_s}\sqrt{\det G}
(2\pi)^p\sqrt{\det \theta}\times \nonumber \\
\times \tr\sum_{m,n,l}
 \bra{n,m,l}A^{mnl}
\sum_{m',n',l'}
\frac{\alpha'}{2}G^{ab}c_0[C_a,[C_b,A^{m'n'l'}]]
 \ket{n',m',l'}=\nonumber \\
=-\frac{2\pi^2 T_{2p-1}}{G_s}\int 
\sqrt{\det G} d^{2p}x
\sum_{m,n,l}
 \bra{n,m,l}A^{mnl}(x) \sum_{m',n',l'}
\frac{\alpha'}{2}G^{ab}c_0\partial_a\partial_bA^{m'n'l'}(x)
\ket{n',m',l'} . \nonumber \\
\end{eqnarray}
Using
\begin{equation}
A(x)^{mnl}=\int d\tilde{k}e^{ikx}A(k)^{mnl} \ ,
\end{equation}
where $\tilde{k}$ is equal to $k(4\pi^2\alpha')^{1/2}$, 
(\ref{e}) is equal to
\begin{eqnarray}\label{imp}
\frac{2\pi^2 T_{2p-1}}{G_s}\sum_{m,n,l,m',n',l'}
\int d\tilde{k}
d\tilde{k}' 
\bra{m,n,l}A^{mnl}(k')
(4\pi^2\alpha')^p\delta (k+k')\frac{\alpha'}{2}\times
\nonumber \\ \times c_0G^{ab}k'_ak_bA(k)^{m'n'l'} \ ,
\ket{m',n',l'} \ , \nonumber \\
\end{eqnarray}
where we have used $ \int \sqrt{\det G}d^{2p}x
e^{ix(k+k')}=(4\pi^2\alpha')^p\delta (k+k')$. From this
definition of the delta function we immediately see that
$\delta (0)=\int \sqrt{\det G} d^{2p}x=V_{26}$.

We see that (\ref{imp}) corresponds to the kinetic term
in the string field theory action for D(2p-1)-brane. More precisely,
after identification $ p^a\sqrt{2\alpha'}=\alpha_0^a$ 
 this expression arises from the 
part of the BRST operator for D(2p-1)-brane proportional to
$\sim c_0g_{IJ}\alpha_0^I\alpha^J_0$  that acts on the string
field in the form
\begin{equation}\label{PsiDp}
\ket{\Psi}=(4\pi^2\alpha')^{p/2}\sum_{n,m,l}
\int d^{2p} \tilde{k} A(k)^{mnl}\ket{m,n,l,k} \ ,
\end{equation}
Using (\ref{Non}) 
we can write the remaining contribution to the zero
mode part of the BRST operator as
\begin{eqnarray}\label{nonzero}
\frac{2\pi^2T_{-1}}{g_s}\tr \bra{\Psi}\frac{\sqrt{2}}{2\pi
\sqrt{\alpha'}}
\sum_{N=-\infty,N\neq 0}^{\infty}
c_Ng_{ab}\alpha^a_{-N}[X^b,\ket{\Psi}]=\nonumber \\
=\frac{2\pi^2T_{2p-1}}{G_s}\sum_{n,m,l,n',m',l',N}
\int d^{2p}\tilde{k}d^{2p}\tilde{k}'
\bra{n,m,l}\sqrt{2\alpha'}c_NA(k)^{nml}
 (4\pi^2\alpha')^p\delta(k+k')
\times \nonumber \\
\times G^{ab}\tilde{\alpha}_{-N,a}k'_bA(k')^{n'm'l'}\ket{n',m',l'}=
\frac{2\pi^2T_{2p-1}}{G_s}\int \bra{\Psi}\sum_{N=-\infty,
N\neq 0}^{\infty}c_{N}G_{ab}\alpha^a_{-N}\alpha_0^b
\ket{\Psi} \ ,
\nonumber \\
\end{eqnarray}
where we have used (\ref{PsiDp}) and we have ommited
tilde on $\alpha$. We see that (\ref{imp}) and
(\ref{nonzero}) give 
the correct contribution to the zero-mode part of
the BRST operator 
\begin{equation}
\frac{1}{2}G_{ab}\alpha_0^a
\alpha_0^b+\sum_{n=-\infty, n\neq 0}^{\infty}c_n
G_{ab}\alpha^a_{-n}\alpha_0^b \ ,
\end{equation}
so that 
 we can claim that $Q^{inst}$ (after appropriate rescaling
$\alpha_n^a$) with
the zero mode part given   above leads to
the BRST operator $Q_{2p-1}$ for D(2p-1)-brane with the noncommutative
world-volume and with the open string parameters given
in (\ref{opp}). 
In order to finish this identification we
must also discuss the interaction part in
 (\ref{Waction}) which has a form
\begin{eqnarray}
\frac{2\pi^2 T_{-1}}{3g_s}\int \tr
\Psi\star \Psi\star \Psi= \frac{2\pi^2T_{2p-1}
}{3G_s}\sqrt{\det G} (2\pi)^p
\sqrt{\det \theta}\int \tr \Psi\star\Psi\star\Psi=\nonumber \\
=\frac{2\pi^2T_{2p-1}}{3G_s}\int \Psi\hat{\star}
\Psi\hat{\star}\Psi \ ,\nonumber \\
\end{eqnarray}
where $\hat{\star}$ is a modified start product that includes
the Moyal star product arising from the noncommutative
nature of the theory.
More precisely, the explicit form of the star product was given
in \cite{LeClair,Gross,Samuel} in terms of some overlap vertices
in the string field theory operator formalism.  The conditions
which these vertices must obey are completely universal for
any background and are completely determined from the form
of the BRST operator. Since we have shown that the resulting
BRST operator corresponds to the noncommutative background,
we could proceed in the same way as in \cite{Sugino,Kawano}
to construct corresponding overlap vertices resulting to the
modification of the start product cited above. It would be certainly
nice to construct overlap vertices for any D-instanton background.
We hope to return to this question in the future. For our purposes
in this paper the abstract definition of the star product
\cite{WittenSFT} is sufficient.

 As a result, we obtain the string field theory
for D(2p-1)-brane in the presence of the background B field
\begin{equation}
S=\frac{2\pi^2T_{2p-1}}{G_s}\int
\left(\frac{1}{2}\Psi\hat{\star} Q_{2p-1}\Psi+
\frac{1}{3}\Psi\hat{\star}\Psi\hat{\star}\Psi\right) \ .
\end{equation}
In this section we have seen that all D-branes of even dimensions
arise from the single D-instanton string field theory with
the modified BRST operator. We have seen that this operator
is correct BRST operator for any background configuration
of D-instantons. In fact, we can regard this BRST operator as 
a particular solution of the pregeometrical string field theory
\cite{Horowitz} which will be seen more preciselly in the next
section. 

We can also generalise this construction to the configuration of $k$ 
D(2p-1)-branes. In this case we take the background configuration
of D-instantons
\begin{equation}\label{Dk}
[X^a,X^b]=1_{k\times k}\otimes i\theta^{ab}1_{N\times N}, \
a, b=1,\dots, 2p \ , X^m=0, \ m=2p+1,\dots,26 \ . 
\end{equation}
It is easy to see that this configuration leads to the non-abelian
$U(k)$ string field theory describing $k$ coincident D(2p-1)-branes. This simply
follows from the decomposition of the string field as
\begin{equation}
\Psi_{IJ}=(\psi_{ab})_{mn} , \
I=m\times N+a , \ J=n\times N+b, \ m,n=0,\dots, k-1 , \
a,b=1,\dots, N, \ N\rightarrow \infty \ .
 \end{equation}
We can easily generalised this solution to solution 
describing  $k$ D(2p-1)-branes
with general transverse positions. We  replace the solution
$X^m=0$ in (\ref{Dk}) with the more general one
\begin{equation}\label{2pk}
X^m=\left(\begin{array}{cccc}
x^m_1\otimes 1_{N\times N} & 0 & \dots & 0 \\
0 & x^m_1  \otimes 1_{N\times N} & \dots & 0 \\
\dots & \dots & \dots & \dots \\
0 & \dots & 0 & x^m_k \otimes 1_{N\times N} \\ \end{array}\right),
\ m=2p+1,\dots,26, \ , N\rightarrow \infty \ .
\end{equation}

\section{Tachyon condensation and noncommutative
string field theory}\label{fourth}

In this section we would like to study the problem of
the emergence  of lower dimensional D-branes from
the string field theory describing space-time filling D25-brane
in the background B-field. As was shown in
\cite{WittenNG}, and for the case of string field theory
in \cite{Sugino,Kawano}, the resulting theory is a noncommutative
one. The precise analysis of the string field theory
was given in the beautiful papers \cite{Sugino,Kawano}
where it was shown that the string field theory 
in the presence of the background B field differs from
the string field theory in the trivial background in
 the modification of the string
field start product which now also incorporates the
Moyal star product of the noncommutative theory. In fact,
we have obtained  this result in the previous section from
slightly different point of view. The
next difference is that all quantities in the string field
theory are calculated with the open string quantities
\cite{WittenNG}.
In other words, the string field action for D25-brane in
the presence of the background B field has  a form
\begin{equation}
S=\frac{2\pi^2T_{25}}{G_s}\int\left(
\frac{1}{2}\Psi\hat{\star}Q\Psi+\frac{1}{3}\Psi
\hat{\star}\Psi\hat{\star}\Psi \right) \ .
\end{equation}
 We will show that with using this action we
will be able to obtain all lower dimensional D-branes of 
even codimensions in the same way as in the
case of the tachyon condensation in the effective field
theory \cite{StromingerA,Gopakumar,
Harvey,Strominger2,HarveyS}.
In order to show this we must say a few words about the
star product in string field theory. In the original Witten's string field theory
\cite{WittenSFT}, the star product was defined as a abstract
operation describing joining two strings which formally
does not depend on the background. On the
other hand, the modified star product depends on
the  background B field through
 the noncommutative parameter. If follows that in
the process of the tachyon condensation the modified
star product $\hat{\star}$  changes since the lower
dimensional D-brane has different noncommutative parameter,
whereas the formal string field star product $\star$ does
not change. For that reason the best thing how to
study the tachyon condensation is in such a way where
we can replace the Moyal star product with the other formulation
of the noncommutative geometry of the world-volume. For
that reason we will transform the string field action into
 the operator formalism \cite{StromingerA,Strominger2}.
Since we will work with the 
 open string parameters 
\cite{WittenNG} given in (\ref{OPM}) we can use the results
presented  in the previous section and express the noncommutative
D25-brane in terms of D-instanton matrix model. 

Now we would like to claim that the emergence of 
any configuration of even dimensional D-branes from 
D25-brane is rather straightforward procedure which does not
need to carry about misterious nothing state
\cite{SenN1,SenN2}. For simplicity,
let us study the emergence of $k$ D2p-branes from 
D25-brane. For that reason we propose the string field $\Phi_0$
that leads to this configuration as follows
\begin{equation}\label{phik}
Q_{2p}^kA=Q_{25}A+\Phi_0\star A-(-1)^AA\star\Phi_0 \ ,
\forall A \ .
\end{equation}
The index $k$ in $Q_{2p}^k$ indices that this is the BRST operator
for any configuration of $k$ D2p-branes that
corresponds to the D-instanton background (\ref{Dk}) and
(\ref{2pk}). 
It is also uderestood the matrix multiplication in (\ref{phik}). 
And finally, we express all quantities in terms of D-instanton
matrix model and in terms of closed string metric and 
coupling constant which is a reflection of the background
independence of the noncommutative theory \cite{WittenNG,Seiberg}.

Since we know that both $Q^k_{2p},Q_{25}$ are nilpotent
operators, (\ref{phik}) leads to the condition that
$\Phi_0$ must be a solution of the string field theory equation
of motion
\begin{equation}
(Q_{25}\Phi_0)_{ij}+(\Phi_0)_{ik}\star(\Phi_0)_{kj}=0 \ .
\end{equation}
We do not give the explicit form of this solution however it
is clear that such a solution should exist from the existence
of the correct BRST operators $Q_{2p}^k,Q_{25}$. 
In fact, we can follow very elegant approach presented in 
\cite{Horowitz} and extend it to the non-abelian case
in the form
\begin{equation}
S=\frac{4\pi^3}{3g_s}\tr \int \Phi\star\Phi\star\Phi \ ,
\ 2\pi^2T_{-1}=4\pi^3 \ ,
\end{equation}
with the equation of motion
\begin{equation}\label{Hor}
\Phi_{ij}\star \Phi_{jk}=0 \ . 
\end{equation}
Following \cite{Horowitz} we can construct for any solution 
of the equation of motion (\ref{Hor})
$\Phi_0$ a (matrix valued) operator $D_{\Phi_0}$
\begin{equation}
(D_{\Phi_0}B)_{ik}=(\Phi_0)_{ij}\star B_{jk}-
(-1)^BB_{ij}\star(\Phi_0)_{jk} \ .
\end{equation}
Then we can  see  (using  axioms
(\ref{ax}) and their generalised form (\ref{nax1}),(\ref{nax2})) that
\begin{eqnarray}
\tr \int D_{\Phi_0}B=
\int [(\Phi_0)_{ij}\star B_{ji}-(-1)^{2B}(\Phi_0)_{ji}\star B_{ij}]=0 \ ,
\nonumber \\ 
\end{eqnarray}
\begin{eqnarray}
(D_{\Phi_0}(A\star B))_{ik}=
(\Phi_0)_{ij}\star A_{jm}\star B_{mk}-(-1)^{A+B}
A_{ij}\star B_{jm}\star(\Phi_0)_{mk}=\nonumber \\
=((\Phi_0)_{ij}\star A_{jm}-(-1)^AA_{ij}\star (\Phi_0)_{jm})
\star B_{mk}
+(-1)^AA_{ij}\star ((\Phi_0)_{jm}\star B_{mk}-
\nonumber \\
-(-1)^B B_{jm}\star (\Phi_0)_{mk})
=(D_{\Phi_0}A)_{ij}\star B_{jk}+(-1)^A
A_{ij}\star ((D_{\Phi_0} B)_{jk} \nonumber \\
\end{eqnarray}
and finally
\begin{eqnarray}
(D_{\Phi_0}^2B)_{im}=
(D_{\Phi_0})_{ij}\left[(\Phi_0)_{jk}\star B_{km}-(-1)^BB_{jk}\star
(\Phi_0)_{km}\right]= \nonumber \\
=((-1)^B\Phi_0\star B\Phi_0-(-1)^B\Phi_0\star B\Phi_0)_{im}=0 \ ,
\nonumber \\ 
\end{eqnarray}
where we have used $(\Phi_0)_{ik}\star(\Phi_0)_{kj}=0$. These
results imply that $D_{\Phi_0}$ is a derivative. 

We will argue that in this way we can construct a
BRST operator in noncommutative theory or equivalently the
BRST operator for infinite number of D-instatons.
In fact, we have implicitely used this construction in the previous
section and the approach given here can serve as futher support
of our proposal.  We 
write the BRST operator as follows
\begin{equation}
(Q_{25}A)_{ij}=Q_{inst}A_{ij}+(Q^0_{25}A)_{ij}=\phi_0\star A_{ij}-
(-1)^AA_{ij}\star \phi_0+(\phi_{25})_{ik}\star A_{kj}
-(-1)^A A_{ij}\star (\phi_{25})_{kj} \ .
\end{equation}
Since we know that $Q_{inst}$ is a correct BRST operator
for the background of $N$ D-instantons sitting in the points
$x^I=0$ 
 then  can be written as 
\begin{equation}\label{fiq0}
Q_{inst}A_{ij}=\phi_0\star A_{ij}-(-1)^AA_{ij}\star \phi_0 \ ,
\end{equation}
where $\phi_0$ can be found as in \cite{Horowitz}.
Using the explicit form of $Q^0_{25}$ 
(\ref{Qinst}) expressed  in
the D-instanton form with the clasical configuration given
in (\ref{non}) we see that it can be written as follows (The
matrix multiplication is underestood) 
\begin{eqnarray}
Q_{25}^0A=Q_{25}^LA+Q_{25}^RA \ , \nonumber \\
Q_{25}^LA=\frac{1}{4\pi^2\alpha'}c_0g_{IJ}[
X^IX^JA-X^IAX^J]+\frac{\sqrt{2}}{2\pi\sqrt{\alpha'}}
\sum_{n=-\infty, n\neq 0}^{\infty}c_n g_{IJ}
\alpha^I_{-n}X^JA \ ,\nonumber \\ 
Q_{25}^RA=\frac{1}{4\pi^2\alpha'}c_0g_{IJ}[
AX^JX^I-X^JAX^I]-\frac{\sqrt{2}}{2\pi\sqrt{\alpha'}}
\sum_{n=-\infty, n\neq 0}^{\infty}c_n g_{IJ}
\alpha^I_{-n}AX^J \ . \nonumber \\ 
\end{eqnarray}
Now we would like to argue that the string field $\phi_{25}$ acts
on any string field $A$ as 
\begin{equation}\label{fiQ}
(\phi_{25})_{ik}\star A_{kj}=(Q^{L}_{25}A)_{ij} \ ,
\end{equation}
which allows us to express the string field in terms
of the zero mode part of the BRST operator $Q_L$.
To support this idea, let us write
\begin{eqnarray}
\int (\phi_{25})_{ij}\star A_{ji}=
(-1)^A\int A_{ji}\star (\phi_{25})_{ij}=\tr \int Q_{25}^LA=\nonumber \\
=\tr \int \left(\frac{1}{4\pi^2\alpha'}c_0g_{IJ}[
X^IX^JA-X^IAX^J]+\frac{\sqrt{2}}{2\pi\sqrt{\alpha'}}
\sum_{n=-\infty, n\neq 0}^{\infty}c_n g_{IJ}
\alpha^I_{-n}X^JA \right)=\nonumber \\
=-\tr\int \left(\frac{1}{4\pi^2\alpha'}c_0g_{IJ}[
AX^JX^I-X^JAX^I]-\frac{\sqrt{2}}{2\pi\sqrt{\alpha'}}
\sum_{n=-\infty, n\neq 0}^{\infty}c_n g_{IJ}
\alpha^I_{-n}AX^J \right)\Rightarrow\nonumber \\ 
(-1)^A\tr\int A\star \phi_{25}=-\tr\int Q_{25}^RA
\Rightarrow (-1)^A A_{ij}\star(\phi_{25})_{jk}
=-(Q_{25}^RA)_{ij}  \nonumber \\
\end{eqnarray}
and consequently
\begin{equation}
(Q_{25}^0A)_{ij}=(\phi_{25})_{ik}\star A_{kj}
-(-1)^AA_{ik}\star(\phi_{25})_{kj}=
(Q_{25}^RA)_{ij}+(Q_{25}^LA)_{ij}=(Q_{25}^0A)_{ij} 
\end{equation}
which we wanted to prove.  
Finally we must also show that $\Phi_0=\phi_0+\phi_{25}$ is 
a solution of the equation of motion for the cubic string field theory
$\Phi\star\Phi=0$. This equation leads to
\begin{equation}
(\Phi_0)_{ij}\star(\Phi_0)_{jk}=
Q_{inst}(\phi_{25})_{ij}+(\phi_{25})_{ik}\star(\phi_{25})_{kj}=0 \ 
\end{equation}
using (\ref{fiq0}).
In other words, $\phi_{0}$ should be a solution of the equation of motion
for D-instanton string field theory. Again, this can be seen from 
 the fact  that $Q_{25}$ is nilpotent operator
(As we have proven in the previous section)  so we have
\begin{equation}
Q_{25}^2=0=Q_{inst}(\phi_{25})_{ij}+(\phi_{25})_{ik}\star
(\phi_{25})_{kj}=0
\end{equation}
so that $\phi_{25}$ is a solution of the equation of motion. 

Using these results it is easy to find string field describing the tachyon
condensation in the noncommutative version of the string field
theory from D25-brane to any lower dimensional configurations
of D2p-branes. We have (in matrix notation)
\begin{eqnarray}\label{2p}
Q_{25}A=\Phi_{25}\star A-(-1)^AA\star \Phi_{25} \ , \nonumber \\
Q_{2p}^kA=\Phi_{2p}\star A-(-1)^AA\star\Phi_{2p} \ , \nonumber \\ 
Q_{2p}^kA=Q_{25}A+\Phi_0\star A-(-1)^AA\star\Phi_0 \Rightarrow 
\Phi_0=\phi_{2p}^0-\phi_{25}^0 \ ,\nonumber \\
\end{eqnarray}
where $\phi_{25},\phi_{2p}^0$ are given in (\ref{fiQ}).
When we rewrite the action for non-commutative D25-brane in 
terms of the matrix model and then we use the upper relation between
the BRST operator and string field $\Phi_{25}$ we can write
the action for non-commutative D25-brane as follows
\begin{equation}\label{2px}
S=\frac{4\pi^3}{3g_s}\tr \int (\Phi_{25}+\Psi)\star
(\Phi_{25}+\Psi)\star(\Phi_{25}+\Psi) \ .
\end{equation}
When we expand around the solution
$\Phi_{0}$  we obtain preciselly
the string field action for the configuration of
$k$ D2p-branes. In other words, when we write
the string field $\Psi$ in (\ref{2px}) as
\begin{equation}
\Psi=\Phi_0+\phi
\end{equation}
and insert it into (\ref{2px}) we get
\begin{equation}
S=\frac{2\pi^2T_{-1}}{3g_s}
\tr\int (\Phi_{2p}+\phi)\star(\Phi_{2p}+\phi)
\star (\Phi_{2p}+\phi)=
\frac{2\pi^2T_{-1}}{g_s}\tr \int
\left(\frac{1}{2}\phi\star Q_{2p}^k\phi+\frac{1}{3}
\phi\star\phi\star\phi \right) \ .
\end{equation}
which is preciselly the string field action for D2p-brane
written in the matrix model formalism.

In this section we have shown that the description
of the string field theory in the noncommutative background
in terms of the generalised matrix string field theory can
very easily describe the emergence of the lower dimensional
D-branes from D25-brane.

\section{Conclusion}\label{sixth}
In this short paper we have tried to present an alternative description
of the Witten's string field theory \cite{WittenSFT} in the presence
of the background B field. We have argued for the existence of
 more general string field action for $N$ D-instantons
 which would have many properties
of the matrix models \cite{BanksM,Banks,Li,Aoki,Ishibashi,Ishibashi1}.
We have seen that matrix description of the string field theory allows
naturally to describe the tachyon condensation to  D-branes of
even dimensions.  
We have made many calculations which should support our proposal. In
  particular, we have shown that the requirament of the
nilpotence of the BRST operator leads to the conclusion
that the background configuration of D-instantions should obey
the equations of motion of the low energy effective theory.

We believe that the matrix string field description can give more 
accurate description of the tachyon condensation. In fact, the importance
of the matrix theory analysis of this problem has been suggested 
previously in \cite{Kraus,LiT}. Of course, in this approach we cannot
describe odd dimensional D-branes which is the same problem as
their description in terms of noncommutative theory. We also cannot
much to say about the tachyon condensation to the closed string vacuum
that is very difficult problem. However, there is now considerable 
progress in its solution \cite{SenN1,SenN2}.

 It would be also very interesting  to try
to extend this analysis to the case of the supersymmetric string field theory.
\\
\\
{\bf Acknowledgements}

We would like to thank Rikard von Unge for very helpful 
discussions. This work was supported by the
Czech Ministry of Education under Contract No.
14310006.


\begin{thebibliography}{40}
\bibitem{SenP}  A. Sen,\emph{"Stable non-BPS bound states of BPS  D-branes,"}
\jhep{9808}{1998}{010},
\hepth{9805029};

\emph{"SO(32) spinors of type I and other solitons 
on brane-antibrane pair,"} 
\jhep{9809}{1998}{023},\hepth{9808141};

\emph{"Type I D-particle and its interactions,"}
\jhep{9810}{1998}{021} \hepth{9809111};

\emph{"Non-BPS states and branes in
string theory,"} \hepth{9904207}, and reference therein.
\bibitem{witen} E. Witten, \emph{"D-branes and 
K-theory,"} \jhep{9812}{019}{1998},  \hepth{9810188}.
\bibitem{Horava} P. Ho\v{r}ava,
\emph{"Type II D-branes, K-Theory and Matrix
Theory,"} 
\atmp{2}{1999}{1373}, \hepth{9812135}.
\bibitem{Olsen} K. Olsen and R. J. Szabo,
\emph{"Constructing D-branes from K-theory,"}
\hepth{9907140}.
\bibitem{witen2} E. Witten,\emph{"Overview Of $K$-theory
Applied To Strings,"} \hepth{0007175}.
\bibitem{Matsuo} Y. Matsuo,
\emph{"Topological Charges of Noncommutative Soliton,"} \hepth{0009002}.
\bibitem{Moore} J. Harvey and G. Moore, \emph{"Noncommutative Tachyons and
K-Theory,"} \hepth{0009030}.
\bibitem{Lerda} A. Lerda and R. Russo,
\emph{"Stable non-BPS D-states in string theory:
a pedagogical review,"} \hepth{9905006}.
\bibitem{Schwarz} J. Schwarz,  
\emph {"TASI Lectures on 
Non-BPS D-Branes Systems,"} \hepth{9908144}.
\bibitem{WittenSFT} E. Witten,
\emph{"Noncommutative geometry and string field theory,"}
\npb{268}{1986}{253}.
\bibitem{SenFT1} A. Sen, 
\emph{"Universality of the tachyon potential,"}
\jhep{9912}{1999}{027}, \hepth{9911116}.
\bibitem{SenFT2} A. Sen and B. Zwiebach,
\emph{"Tachyon Condensation in String Field
Theory,"}
\jhep{0003}{002}{2000}, \hepth{9912249}.
\bibitem{BerkovitzFT1} N. Berkovits,\emph{"The Tachyon Potential in Open Neveu-
Schwarz String Field Theory,"} 
\jhep{0004}{022}{2000}, \hepth{0001084}.
\bibitem{HarveyFT} J. A. Harvey and P. Kraus,
\emph{"D-Branes as Lumps in Bosonic Open
String Field Theory,"}
\jhep{0004}{012}{2000}, \hepth{0002117}.
\bibitem{SenFT3} N. Berkovits, A. Sen and B. Zwiebach,
\emph{"Tachyon Condensation in Superstring Field
Theory,"} \hepth{0002211}.
\bibitem{TaylorFT} N. Moeller and W. Taylor,
\emph{"Level truncation and the tachyon in open
bosonic string field theory,"}
\npb{563}{105}{2000}, \hepth{0002237}.
\bibitem{KochFT} R. de Mello Koch, A.
Jevicki, M. Mihailescu and R. Tatar,
\emph{"Lumps and p-branes in open string field theory,"}
\plb{482}{249}{2000}, \hepth{0003031}.
\bibitem{Desmet} P. J. De Smet and 
J. Raeymaekers,
\emph{"Level-four approximation to the tachyon
potential in superstring field theory,"}
\jhep{0005}{051}{2000}, \hepth{0003220}.
\bibitem{NaqviFT} A. Iqbal and A. Naqvi,
\emph{"Tachyon Condensation On A Non-BPS
D-Brane,"} \hepth{0004015}.
\bibitem{SenFT4} N. Moeller, A. Sen and
B. Zwiebach, \emph{"D-branes as Tachyon Lumps
in String Field Theory,"} 
\jhep{0008}{039}{2000}, \hepth{0005036}.
\bibitem{DavidFT} J. R. David,
\emph{"$U(1)$ gauge invariance from open 
string field theory,"} \hepth{0005085}.
\bibitem{WittenFT} E. Witten,
\emph{"Noncommutative Tachyons And String Field
Theory,"} \hepth{0006071}.
\bibitem{RasteliFT} L. Rasteli and B. Zwiebach,
\emph{"Tachyon Potentials, Star Products and Universality,"}
\hepth{0006240}.
\bibitem{SenFT5} A. Sen and B. Zwiebach,
\emph{"Large Marginal Deformations in String
Field Theory,"} \hepth{0007153}.
\bibitem{TaylorFT2} W. Taylor,
\emph{"Mass generation from tachyon condensation
for vector fields on D-brane,"} \hepth{0008033}.
\bibitem{KochFT2} R. de Mello Koch and
J. P. Rodrigues, \emph{"Lumps in level truncated
open string field theory,"} \hepth{0008053}.
\bibitem{NaqviFT2} A. Iqbal and A. Naqvi,
\emph{"An Marginal Deformations in Superstring
Field Theory,"} \hepth{0008127}.
\bibitem{Kostelecky} A. Kostelecky and R. Potting,
\emph{"Analytical construction of a nonperturbative
vacuum for the open bosonic string,"}
\hepth{0008252}.
\bibitem{Schnabl} M. Schnabl,
\emph{"String field theory at large B-field 
and noncommutative geometry,"} \hepth{0010034}.
\bibitem{SenN1} L. Rastelli, A. Sen and B. Zwiebach,
\emph{"String Field Theory Around the Tachyon Vacuum,"}
\hepth{0012251}.
\bibitem{Ohmori} K. Ohmori,
\emph{"A Review on Tachyon Condensation in Open String
Field Theories,"} \hepth{0102085}.
\bibitem{Hata} H. Hata and S. Teraguchi,
\emph{"Test of the Absence of Kinetic Terms around the
Tachyon Vacuum in Cubic String Field Theory,"} \hepth{0101162}.
\bibitem{SenN2} L. Rastelli, A. Sen and B. Zwiebach,
\emph{"Classical Solutions in String Field Theory Around the
Tachyon Vacuum,"} \hepth{0102112}.
\bibitem{TaylorN}  I. Ellwood and W. Taylor,
\emph{"Open string field theory without open strings,"}
\hepth{0103085}.
\bibitem{Feng} B. Feng, Y. He and N. Moeller,
\emph{"Testing the Uniqueness of the Open String Field Theory Vacuum,"}
\hepth{0103103}.
\bibitem{WittenBT} E. Witten,
\emph{"On background independent open string
field theory,"} \prd{36}{5467}{1992}, \hepth{9208027}.
\bibitem{WittenBT1} E. Witten,
\emph{"Some computations in background 
independent off-shell string theory,"}
\prd{47}{3405}{1993}, \hepth{9210065}.
\bibitem{WittenBT2} K. Li and E. Witten,
\emph{"Role of short distance behaviour in
off-shell open string field theory,"} 
\prd{48}{853}{1993}, \hepth{9303067}.
\bibitem{ShatasviliBT} S. L. Shatashvili,
\emph{"Comment on the background independent 
open string theory,"} \plb{311}{83}{1993},
\hepth{9303143}.
\bibitem{ShatasviliBT1} S. L. Shatashvili,
\emph{"On the problems with background independence
in string theory,"} \hepth{9311177}.
\bibitem{ShatasviliBT2} A. A. Gerasimov and
S. L. Shatashvili, 
\emph{"On exact tachyon potential in
open string field theory,"}, \jhep{0010}{034}{2000},
 \hepth{0009103}.
\bibitem{MooreBT} D. Kutasov, M. Marino and
G. Moore, \emph{"Some exact results on
tachyon condensation in string field theory,"}
\hepth{0009148}.
\bibitem{SenBT} D. Ghoshal and
A. Sen, \emph{"Normalisation of the Background 
Independent Open String Field Theory Action,"}
\hepth{0009191}.
\bibitem{Cornalba} L. Cornalba,
\emph{"Tachyon Condensation in Large Magnetic
Fields with Background Independent String Field
Theory,"} \hepth{0010021}.
\bibitem{Okuyama} K. Okuyama,
\emph{"Noncommutative Tachyon from Background
Independent Open String Field Theory,"} \hepth{0010028}.
\bibitem{MooreBT2} D. Kutasov, 
M. Marino and G. Moore,
\emph{"Remarks on Tachyon Condensation in
Superstring Field Theory,"} \hepth{0010108}.
\bibitem{DasguptaBT1} S. Dasgupta  and T. Dasgupta,
\emph{"Renormalisation Group Analysis of Tachyon
Condensation,"} \hepth{0010247}.
\bibitem{ShatasviliBT3}
A. A. Gerasimov and S. L. Shatashvili,
\emph{"String Higgsy Mechanism and the Fate
of Open Strings,"} \hepth{0011009}.
\bibitem{KrausBT} P. Kraus and F. Larsen,
\emph{"Boundary String Field Theory of the
$D\overline{D}$-System,"} \hepth{0012198}.
\bibitem{TakayanagiBT} T. Takanayagi, S. Terashima
and T. Uesugi, \emph{"Brane-Antibrane Action
from Boundary String Field Theory,"} \hepth{0012210}.
\bibitem{WittenNG} N. Seiberg and E. Witten,
\emph{"String Theory and Noncommutative Geometry,"}
\jhep{9909}{032}{1999}, \hepth{9908142}.
\bibitem{StromingerA} R. Gopakumar, S. Minwalla and 
A. Strominger, \emph{"Noncommutative solitons,"}
\jhep{0006}{022}{2000}, \hepth{0003160}.
\bibitem{Gopakumar} K. Dasgupta, S. Mukhi and
G. Rajesh, \emph{"Noncommutative Tachyons,"}
\jhep{0006}{02}{2000}, \hepth{0005006}.
\bibitem{Harvey} J. A. Harvey, P. Kraus, F. Larsen
and E. J. Martinec, \emph{"D-branes and strings as
noncommutative solitons,"}
\jhep{0007}{042}{2000}, \hepth{0005031}.
\bibitem{Seiberg} N. Seiberg, 
\emph{"A Note on Background Independence in
Noncommutative Gauge Theories, Matrix Model
and Tachyon Condensation,"} \hepth{0008013}.
\bibitem{Harvey2} J. A. Harvey, P. Kraus and F. Larsen,
\emph{"Tensionless Branes and Discrete Gauge 
Symmetry,"} \hepth{0008064}.
\bibitem{Strominger2} R. Gopakumar, S. Minwalla and
S. Strominger, 
\emph{"Symmetry Restoration and Tachyon Condensation
in Open String Theory,"} \hepth{0007226}.
\bibitem{Rey} G. Mandal and S. J. Rey,
\emph{"A note on D-Branes of Odd Codimensions
from Noncommutative Tachyons,"} \hepth{0008214}.
\bibitem{SenTP} A. Sen,\emph{"Some Issues in
Non-Commutative Tachyon Condensation,"}
\hepth{0009038}.
\bibitem{Mukhi} S. Mukhi and N. V. Suryanarayana,
\emph{"Chern-Simons Terms on Noncommutative 
Branes,"} \hepth{0009101}.
\bibitem{Strominger3} M. Aganagic, R. Gopakumar,
S. Minwalla and A. Strominger,
\emph{"Unstable Solitons in Noncommutative 
Gauge Theory,"} \hepth{0009142}.
\bibitem{HarveyS} J. A. Harvey, P. Kraus and
F. Larsen, \emph{"Exact noncommutative solitons,"}
\hepth{0010060}.
\bibitem{Larsen} F. Larsen, \emph{"Fundamental Strings as
Noncommutative Solitons,"} \hepth{0010181}.
\bibitem{KlusonM} J. Kluso\v{n},
\emph{"D-Branes from N Non-BPS D0-branes,"}
\hepth{0009189}.
\bibitem{Kraus} P. Kraus, A. Rajaraman and
S. Shenker, \emph{"Tachyon Condensation in Noncommutative
Gauge Theory,"} \hepth{0010016}.
\bibitem{LiT} M. Li,
\emph{"Note on Noncommutative Tachyon in Matrix
Models,"} \hepth{0010058}.
\bibitem{Mandal} C. Mandal and S. R. Wadia,
\emph{"Matrix model, Noncommutative Gauge Theory and
the Tachyon Potential,"} \hepth{0011094}.
\bibitem{Alwis} S. P. de Alwis and A. T. Flourhoy,
\emph{"Some Issues in Noncommutative Solitions as
D-branes,"} \hepth{0011223}.
\bibitem{LeClair} A. LeClair, M. E. Peskin and C. R. 
Preitschopf,
\emph{"String Field Theory on The Conformal Plane 1.
Kinematical Principles,"} \npb{317}{411}{1989};
\emph{"String Field Theory on The Conformal Plane 2. 
Generalised Gluing,"} \npb{317}{464}{1989}.
\bibitem{Sugino} F. Sugino,
\emph{"Witten's opens string field theory in  constant
B-field background,"} \jhep{003}{017}{2000}, \hepth{9912254}.
\bibitem{Kawano} T. Kawano and T. Takahashi,
\emph{"Open string field theory on noncommutative
space,"} \hepth{9912274}.
\bibitem{Gross} D. J. Gross and A. Jevicki,
\emph{"Operator Formulation of Interacting String
Field Theory I, II,"}
\npb{283}{1}{1987}, \npb{287}{225}{1987}.
\bibitem{Samuel} S. Samuel,
\emph{"The Physical and Ghost Vertices In
Witten's  String Field Theory,"} \plb{181}{255}{1986}.
\bibitem{BanksM} T. Banks, W. Fischer, S. Shenker
and L. Susskind, 
\emph{"M Theory as a Matrix Model: A Conjecture,"}
\prd{55}{1997}{5112}, \hepth{9610043}.
\bibitem{Banks} T. Banks  N. Seiberg and
S. Shenker, \emph{"Branes from Matrices,"}
\npb{497}{1997}{41}, \hepth{9612157}.
\bibitem{Li} M. Li,
\emph{"Strings from IIB Matrices,"} \npb{499}{1997}{149},
\hepth{9612222}.
\bibitem{Aoki} H. Aoki, N. Ishibashi, S. Iso, H. Kawai,
Y. Kitazawa and T. Tada,
\emph{"Noncommutative Yang-Mills in IIB Matrix Model,"}
\npb{565}{2000}{176}, \hepth{9908141}.
\bibitem{Ishibashi} N. Ishibashi,
\emph{"A relation between commutative and noncommutative
descriptions of D-branes,"} \hepth{9909176}.
\bibitem{Ishibashi1} N. Ishibashi, H. Kawa and Y. Kitizawa,
\emph{"Wilson Loops in Noncommutative Yang-Mills,"}
\hepth{9910004}.
\bibitem{Horowitz} G. T. Horowitz, J. Lykken, R. Rohm
and A. Strominger,
\emph{"Purely Cubic Action for String Field Theory,"}
\prl{57}{2}{1986}.
\bibitem{ZwiebachT1} J. A. Minahan and B. Zweibach,
\emph{"Field theory models for tachyon and gauge
field string dynamics,"}
\jhep{0009}{029}{2000},  \hepth{0008231}.
\bibitem{ZwiebachT2} J. A. Minahan and
B. Zwiebach, \emph{"Effective tachyon dynamics in
superstring theory,"} \hepth{0009246}.
\bibitem{SenNG} A. Sen, \emph{"Fundamental Strings
in Open String Theory at the Tachyonic Vacuum,"} 
\hepth{0010240}. 
\bibitem{WittenBook} M. B. Green, 
J. Schwarz and E. Witten,
\emph{"Superstring theory,"} Vol. 1, Cambridge University
Press, 1987.
\bibitem{ZwiebachAB1} B. Zwiebach,
\emph{"Oriented open-closed string theory revised,"}
\ap{267}{1988}{193}, \hepth{9705241}.
\bibitem{ZwiebachAB2} M. Gaberdiel and
B. Zwiebach, \emph{"Tensor constructions of open string
theories I:Foundations,"} \npb{505}{1997}{569}, \hepth{9705038}.
\bibitem{Lazaraiu1} C. I. Lazaroiu, \emph{"On the structure of
open-closed topological field theory in two dimensions,"}
\hepth{0010269}.
\bibitem{Lazaraiu2} C. I. Lazaroiu, \emph{"Generalised complexes
and string field theory,"} \hepth{0102122}.
\bibitem{Lazaraiu3} C. I. Lazaroiu, \emph{"Unitarity,
D-brane dynamics and D-brane categories,"} \hepth{0102183}.
\end{thebibliography}
\end{document}